# A TWO INTERMEDIATES AUDIO STEGANOGRAPHY TECHNIQUE

Youssef Bassil
LACSC – Lebanese Association for Computational Sciences
Registered under No. 957, 2011, Beirut, Lebanon
youssef.bassil@lacsc.org

## ABSTRACT

On the rise of the Internet, digital data became openly public which has driven IT industries to pay special consideration to data confidentiality. At present, two main techniques are being used: Cryptography and Steganography. In effect, cryptography garbles a secret message turning it into a meaningless form; while, steganography hides the very existence of the message by embedding it into an intermediate such as a computer file. In fact, in audio steganography, this computer file is a digital audio file in which secret data are concealed, predominantly, into the bits that make up its audio samples. This paper proposes a novel steganography technique for hiding digital data into uncompressed audio files using a randomized algorithm and a context-free grammar coupled with a lexicon of words. Furthermore, the proposed technique uses two intermediates to transmit the secret data between communicating parties: The first intermediate is an audio file whose audio samples, which are selected randomly, are used to conceal the secret data; whereas, the second intermediate is a grammatically correct English text that is generated at runtime using a context-free grammar and it encodes the location of the random audio samples in the audio file. The proposed technique is stealthy and irrecoverable in a sense that it is difficult for unauthorized third parties to detect the presence of and recover the secret data. Experiments conducted showed how the covering and the uncovering processes of the proposed technique work. As future work, a semantic analyzer is to be developed so as to make the intermediate text not only grammatically correct but also semantically plausible.

## 1. INTRODUCTION

With the advent of the Internet, computer users started to distribute, share, and transmit their private data online in a complete overt manner. As a result, securing these data became a critical issue for everyone. Steganography and cryptography are two security methods that provide data confidentiality [1]. In essence, cryptography ciphers the secret message so that it cannot be read by eavesdroppers; while, steganography conceals the secret message into a computer file so that it cannot be seen by eavesdroppers [2]. To a third party, an encrypted message would right away imply a secret communication. However, a concealed message would not draw any attention and therefore would not raise suspicions that a secret communication is taking place. For this reason, steganography is often regarded as a stealthy method for transmitting sensitive data into total secrecy across public channels in such a way that no one, apart for the communicating parties namely the original sender and the intended receiver, can know about the existence of the communication [3]. Basically, audio steganography is a type of digital steganography that hides digital data into digital audio files such WAV, MP3, and WMA files [4]. Audio steganography takes advantage of the Human Auditory System (HAS) which cannot hear the slight variation of audio frequencies at the high frequency side of the audible spectrum; and thus, audio steganography can exploit and use this type of frequencies to hide secret data without damaging the quality of the audio file or changing its size [5].

This paper proposes a novel randomized steganography algorithm for hiding digital data into uncompressed audio files using two carrier intermediates to deliver the secret data. The first intermediate is a carrier audio file holding the secret data inside the LSBs of its audio samples which are selected randomly. The second intermediate is a grammatically correct English paragraph made up of several English sentences pointing to the location of the random carrier audio samples, that is, the location of the secret data in the carrier audio file. Basically, the second intermediate is dynamically generated during the hiding process using a context-free grammar of the English language and a lexicon of English words randomly categorized in 10 categories representing the 10 digits of the decimal system. These digits are used to generate the values all possible locations of the carrier audio samples. As the proposed algorithm is randomized, it has such advantages as being irrecoverable in a sense that it is difficult for anyone apart from the original communicating parties to detect the presence of the secret data into the carrier audio file and recover eventually these data.

## 2. AUDIO STEGANOGRAPHY & METHODS OF HIDING

Fundamentally, audio steganography is the art and science of hiding digital data such as text messages, documents, and binary files into audio files such as WAV, MP3, and RM files. The output audio file is called the carrier file and is the only intermediate to be sent to the receiver. Characteristically, anyone who is taping the communication wire would not notice anything suspicious being transmitted, except for a normal audio file [6]. This property of steganography is called imperceptibility and it refers to the fact that no one apart from the original sender and the intended receiver can suspect the presence of secret data into the carrier file being communicated. Steganography can be achieved by means of three types of techniques: injection, substitution, and generation [7].

**Injection:** The injection technique implants the data to hide in the insignificant part of the carrier file, which is normally ignored by operating systems and software applications. For example, most computer files comprise what so called an end-of-file marker or EOF for short, which indicates that no more data can be read from a data source. Likewise, an executable file usually ends with an EOF marking the end of binary instructions. Another example is the PDF file which ends with an EOF indicating to the reader application that no more pages are to be fetched and that the file has ended. Steganography by injection exploits the EOF section and injects secret data after the EOF marker which eventually has no side effect on the carrier file and is often disregarded by the execution environment.

**Substitution:** The substitution technique substitutes the insignificant bits in the carrier file with the bits of the data to hide. Insignificant bits are those bits that can be modified without damaging the quality or destroying the integrity of the carrier file. For example, in audio files, every unit of sound is made up of a sequence of bits. If the least significant bit of this sequence is modified, its impact is minimal on the perceptible sound so much so that the human ear cannot tell the difference between the original version and the altered one. This technique takes advantage of the limited capabilities of the human auditory system (HAS) which cannot recognize two sounds that are slightly not alike.

**Generation:** The generation technique reads the data to hide and generates out of them a new set of data. It is a dynamic method of creating a carrier file based on the information contained in the data to hide. For example, one generation technique will take the message to hide and turn its characters into matching audio frequencies that can ultimately make up an audio file.

## 3. STATE-OF-THE-ART IN AUDIO STEGANOGRAPHY

The popularity and the abundance of audio files make them qualified to convey secret information. As a result, many researchers started to investigate how audio signals and audio properties can be used in the domain of information hiding [8]. Several approaches were conceived, the most popular ones are Least Significant Bit [9], Echo hiding [10], Hiding in Silence Interval [11], Phase Coding [12], Amplitude Coding [13], Spread Spectrum [14], and Discrete Wave Transform [15]:

**Least Significant Bit (LSB):** Basically, the LSB technique is based on embedding each bit from the data to hide into the rightmost bits of every audio sample of the carrier audio file. The LSB technique takes advantage of the HAS which cannot hear the slight variation of audio frequencies at the high frequency side of the audible spectrum. The LSB technique allows high embedding rate without degrading the quality of the audio file. Furthermore, it is relatively effective and easy to implement. However, it main drawback is that the secret data are concealed in a very predictable way, making them easy to be recovered by attackers.

**Echo hiding:** In this technique, the secret data are embedded into the audio signals as a short acoustic echo. In fact, an echo is a replication of sound, however, received by the listener some time after the original sound. As the echo is audible, its amplitude must be decreased so that it becomes imperceptible. In order to hide data, bits whose values are 0 are represented by an echo delayed 1ms; bits whose values are 0 are represented by an echo delayed 2ms. The limitation of echo hiding technique is the low hiding capacity as it would be computationally intensive to insert echo for every bit to hide.

**Hiding in Silence Interval:** This technique inserts a silence interval in the original audio signal to embed the secret data. The values that represent the length of the silence intervals are decreased by some value such as $0<value<2^n$ where n is the number of bits required to represent an element from the data to hide. The carrier audio file is then sent to the receiver having the new lengths for its silence intervals. Recovering the data is done via *mod(altered_length; $2^n$)*.

**Phase Coding:** This technique substitutes the phase of an audio sample with a reference phase that expresses the secret data. The remaining samples are attuned so as to preserve the relative level between different audio samples. Algorithmically, the audio signal is divided into smaller samples whose size is equal to the size of the message to hide. Then, DFT (Discrete Fourier Transform) is applied to generate a matrix of phases. Then, the phase alteration between the contagious samples is calculated. Afterwards, the absolute phases are changed so that to embed the secret data in phase vector of the first audio sample. Finally, the audio signal is rebuilt by computing the DFT using the new generated phase matrix and the original matrix. Consequently, the sound samples are grouped together yielding to a carrier audio signal that encodes the secret data into it.

**Amplitude Coding:** This technique conceals secret data in the magnitude speech spectrum while not distorting the carrier audio signal. It is based on searching for safe spectral regions in the signal whose magnitude speech spectrum is below a certain value. Besides, the carrier locations are selected based on how much they can badly affect the audio signal.

**Spread Spectrum:** This technique scatters the secret data over the frequency spectrum of the audio file using a specific code independent of the actual signal. Basically, secret data are multiplied by a code known to the communicating parties only, and then embedded in the carrier audio file. The advantage of Spread Spectrum method is its speed in covering data; however, its drawback is that it introduces noise and distortions to the audio file.

**Discrete Wave Transform:** In this technique secret data are embedded in the least significant bits of the wavelet coefficients of the audio signals. Often, secret data are chosen to be hidden in the integer wavelet coefficients and not in silent sections of the audio signal so as to promote the imperceptibility of the audio file. The disadvantage of Discrete Wave Transform is that secret data can be lost during the recovering process as this technique is not that accurate.

## 4. THE PROPOSED TECHNIQUE

This paper proposes a novel steganography technique for hiding any form of digital data into uncompressed digital audio files in a random manner. It uses two intermediates to convey the secret data. An uncompressed audio file acting as a carrier file holding the secret data inside the LSBs of its audio samples, and an English text made up of several well-structured and syntactically correct English sentences pointing to the location of the carrier audio samples, that is, the location of the secret data in the carrier audio file. Algorithmically, the proposed technique is based on a randomized algorithm to randomly select the carrier audio samples into which the secret data are be concealed, and on a context-free grammar of the English language to generate correct English sentences that encode the location of the random carrier audio samples in the carrier audio file. The carrier audio file is processed as a 2D map with coordinates x and y (x, y) representing the different samples' locations. The employed context-free grammar uses a lexicon of English words randomly categorized in 10 categories representing the 10 digits of the decimal numeral system (0…9). These digits are used to generate all possible coordinate values for the carrier audio samples.

### 4.1 Basics of the Proposed Technique

The proposed technique is designed to work on 16-bit uncompressed digital audio files such as WAV files. Basically, the samples of a 16-bit WAV file are each composed of two bytes which together represent the amplitude of the audio sample [16]. The proposed technique hides the secret data into the three LSBs of each of these audio samples; thus, the hiding capacity is equal to 3 bits out of 16 bits or 18% of the total size of the carrier audio file (3/16=0.18=18%). The carrier audio file is manipulated as a 2D map composed of a finite set of points along with their coordinates. These coordinates effectively indicate the locations of the audio samples in the carrier file itself. Figure 1 depicts the samples of a WAV audio file as a 2D map as regarded by the proposed technique.

|  | 16-bit | 16-bit | 16-bit | 16-bit |
|---|---|---|---|---|
| 16-bit | audio sample (0,0) | audio sample (0,1) | audio sample (0,2) | audio sample (0,n) |
| 16-bit | audio sample (1,0) | audio sample (1,1) | audio sample (1,2) | audio sample (1,n) |
| 16-bit | audio sample (2,0) | audio sample (2,1) | audio sample (2,2) | audio sample (2,n) |
| 16-bit | audio sample (m,0) | audio sample (m,1) | audio sample (m,2) | audio sample (m,n) |

**Figure 1:** Carrier Audio File as a 2D Map with Coordinates

In effect, the proposed technique does not hide the secret data sequentially into the carrier audio samples; instead, it randomly selects the audio samples to carry in the secret data. Thus, shuffling and dispersing the secret data over the carrier audio file in a random manner. The actual positions of these randomly selected carrier samples, which interchangeably are their coordinates, are mimicked by an English text composed of English sentences that are grammatically well structured. The English text is generated using a context-free grammar and a lexicon of words organized into 10 categories from category 0 till category 9. Words that constitute the English sentences are selected from these categories based on the coordinates of the carrier audio samples. For instance, coordinates (019, 421) are encoded by selecting a word from category 0, a word from category 1, a word from category 9, a word from category 4, a word from category 2, and a word from category 1 respectively. Actually, every coordinate (i.e. audio sample location) is represented by an English sentence. As a result, multiple audio samples would result in multiple sentences which eventually result in a complete English text. The context-free grammar is essential in order to generate a grammatically correct text, for instance, generating a sentence composed of a determinant, followed by a noun, followed by verb, followed by a preposition. It is worth noting that no common words exit between the categories of the lexicon as it would be later impossible to recover the real coordinates of the carrier audio samples out the English text. Finally, together, the carrier audio file and the generated English text are sent to the receiver who has to use the same algorithm, the same grammar, and the same lexicon to recover the different locations of the carrier audio samples and consequently the secret data.

### 4.2 The Context-Free Grammar and the Lexicon

A context-free grammar or CFG is a mathematical system for modeling the structure of languages such as natural languages like English, French and Arabic, or computer programming languages like C++, Java, and C# [17]. Its formalism was originally set by Chomsky [18] and Backus [19], independently of each other. The one of Backus is known as the Backus-Naur Form or BNF for short [20]. In essence, a context-free grammar consists of a set of rules known as production rules that specify how symbols and words of a language can be arranged and grouped together. An example of a rule would specify that in the English language a verb phrase "VP" must always start with a verb then followed by a noun phrase "NP". Another rule would state that a noun phrase "NP" can start with a proper noun "ProperNoun" or a determinant "Det". Following is a sample CFG for a hypothetical language called L, that is a subset of the English language.

*NP → Det Nominal*
*NP → ProperNoun*
*Nominal → Noun | Nominal Noun*
*Det → an*
*Det → the*
*Noun → apple*

The symbols that are used in a CFG are divided into two classes: Terminals that correspond to unbreakable words in the language such as "apple", "the", and "an"; and non-terminals which are variables that can be replaced by terminals and other non-terminal variables to derive and produce sentences of the language. Terminals are usually provided by a lexicon of words; whereas, the non-terminals and the production rules are part of the parsing algorithm. Parsing is the process of deriving sentences for the language using the production rules of the CFG and the lexicon [21]. Parsing can also be used to confirm that the structure of a sentence complies with the CFG of the language.

The proposed steganography technique uses a mini-version CFG of the English language and a lexicon of predefined words to generate correct English sentences that can encode the coordinates of the carrier audio samples. Figure 2 outlines the production rules of the CFG used by the proposed technique.

$S \rightarrow NP\ VP$
$S \rightarrow VP$
$NP \rightarrow Pronoun$
$NP \rightarrow Proper\text{-}Noun$
$NP \rightarrow Det\ Nominal$
$Nominal \rightarrow Noun$
$Nominal \rightarrow Nominal\ Noun$
$Nominal \rightarrow Nominal\ PP$
$VP \rightarrow Verb$
$VP \rightarrow Verb\ NP$
$VP \rightarrow Verb\ NP\ PP$
$VP \rightarrow Verb\ PP$
$VP \rightarrow VP\ PP$
$PP \rightarrow Preposition\ NP$

**Figure 2:** CFG of the Proposed Technique

On the other hand, the lexicon of the CFG which is outlined in Table 1 is organized into 10 categories each of which contains a set of terminal words that belong to the English language. It is worth mentioning that no common words exist between the categories of the lexicon as it would be later impossible to recover the real coordinates of the carrier audio samples out of the English text.

**Table 1:** Lexicon of the Proposed CFG

| Category 0 | Category 1 |
|---|---|
| Det → this \| that \| … <br> Pronoun → I \| they \| us \| … <br> Preposition → from \| across \| about \| … <br> Noun → door \| car \| memory \| … <br> Verb → play \| eat \| walk \| … <br> Proper-Noun → California \| John \| Intel \| … | Det → those \| an \| … <br> Pronoun → he \| she \| me \| … <br> Preposition → on \| above \| through \| … <br> Noun → tree \| school \| board \| … <br> Verb → study \| dance \| climb \| … <br> Proper-Noun → Texas \| NBA \| … |
| Category 2 | … Category 9 |
| Det → a \| these \| … <br> Pronoun → you \| it \| their \| … <br> Preposition → to \| towards \| along \| … <br> Noun → girl \| university \| roof \| … <br> Verb → sleep \| like \| enroll \| … <br> Proper-Noun → Ohio \| George \| Mike \| … | Det → the \| … <br> Pronoun → we \| ours \| his \| … <br> Preposition → among \| at \| before \| … <br> Noun → floor \| book \| pen \| … <br> Verb → move \| walk \| write \| … <br> Proper-Noun → Harvard \| Tony \| … |

Every word in the above categories uniquely encodes the category to which it belongs. For instance, the word "this" encodes the digit "0" exclusively because it belongs to category 0. The word "Harvard" encodes the digit "9" exclusively because it belongs to category 9, and so forth. It is not mandatory, in practice, that the lexicon of Table 1 is used exactly as is; the communicating parties can compile their own lexicon and use it mutually for the covering as well as the uncovering operation.

**4.3 The Proposed Algorithm**

The proposed algorithm consists of several steps needed to be executed in sequence to hide some input secret data into an audio WAV file.

1. The secret data are preprocessed so that they become suitable for storage inside the carrier audio file.
    a. The secret data, no matter their types and formats – whether text, documents, or executables, are converted into a binary form resulting into a string of bits denoted by D={$b_0, b_1, b_2, b_3,…b_{n-1}$} where $b_i$ is a single bit composing the secret data and $n$ is the total number of bits.
    b. The string of bits D is organized into chunks of 3 bits, such as D=Chunks={ $C_0[b_0, b_1, b_2]$, $C_1[b_3, b_4, b_5]$, $C_2[b_6, b_7, b_8]$,…$C_{m-1}[b_{n-3}, b_{n-2}, b_{n-1}]$ }, where $C_j$ is a particular 3-bit chunk, $m$ is the total number of chunks, and $n$ is the total number of bits making up the secret data.
2. A set of carrier audio samples is randomly selected from the carrier audio file in which every random selected sample is denoted by $S_t(x_t, y_t)$ where $x$ and $y$ are the coordinates of sample S, and $t$ is the index of S.
3. The chunks of the secret data that were created in step 1.b are embedded into the randomly selected audio samples of step 2.
    a. As the carrier audio file is sampled as 16-bit, every audio sample would be of length 16 bits. For this reason, every chunk $C_j$ is stored in the three LSBs of every audio sample such as $S_t$={ *thirteenMSBs*($S_t$) + $C_j$ }, where S is a randomly selected audio sample, and $t$ is the index of S. Furthermore, *thirteenMSBs*(S) is a function that returns the original thirteen most significant bits of the audio sample. The "+" operator concatenates the original thirteen MSBs of the audio sample with the 3 bits of a particular chunk, making the total number of bits in a given audio sample equals to 16 bits. In effect, the first thirteen bits are the original thirteen MSBs of the audio sample and the three LSBs are a particular chunk from the secret data.
4. The coordinates of the randomly selected audio samples of step 2 are encoded into English sentences each of which is denoted by $V_k$.
    a. Every single digit of the coordinates $x$ and $y$ such as $x$={$d_0,d_1,d_2,d_{r-1}$} and $y$={$d_0,d_1,d_2,d_{r-1}$} is mapped into a category number of the lexicon. For instance, P(29, 01) is encoded using the CFG of Figure 2 and the lexicon of Table 1 as "a book from school". In that, "a" is a determinant from category 2, "book" is a noun from category 9, "from" is a preposition from category 0, and "school" is a noun from category 1.
    b. Step 4.a is repeated for all selected audio samples. Eventually, the number of all generated English sentences would be equal to the number of all randomly selected audio samples.
5. The final output comprises two intermediates. The first one is the carrier audio file which houses the secret data into its randomly selected samples such as WAV={$S_0,S_1,S_2,S_{t-1}$}, where S is a carrier audio sample and $t$ is the total number of carrier audio samples; while, the second one is an English text made up of grammatically correct English sentences such as T={$V_0,V_1,V_2,V_{t-1}$}, where V is an English sentence and $t$ is the total number of English sentences which is also equal to the total number of carrier audio samples. Both, the carrier audio file and the English text are to be sent to the receiver, not necessarily at the same time, however, they should be both present when the receiver is uncovering the secret data.

## 5. EXPERIMENTS & RESULTS

In the experiments a real example was tested to illustrate how the proposed steganography technique works. It involved using the different steps of the algorithm to cover a secret text message into a 16-bit WAV carrier audio file. The secret data to hide is a text message denoted by D="kill joe".

1. The secret message is preprocessed and converted into a binary form.
   a. The secret message in ASCII format can be denoted as $D_{ASCII}$={ k=1101011  i=1101001  l=1101100  l=1101100  space=0100000  j=1101010  o=1101111  e=1100101 }
   b. Chunks of 3 bits are generated out of D and are denoted by Chunks={ $C_0$[110], $C_1$[101], $C_2$[111], $C_3$[010], $C_4$[011], $C_5$[101], $C_6$[100], $C_7$[110], $C_8$[110], $C_9$[001], $C_{10}$[000], $C_{11}$[001], $C_{12}$[101], $C_{13}$[010], $C_{14}$[110], $C_{15}$[111], $C_{16}$[111], $C_{17}$[001], $C_{18}$[01]. The total number of chunks is 19; and therefore, 19 audio samples are required to store these chunks.
2. Nineteen audio samples are randomly selected. Their coordinates are respectively: $S_0$(206, 318), $S_1$(407, 192), $S_2$(321, 129), $S_3$(709, 015), $S_4$(501, 000), $S_5$(712, 200), $S_6$(309, 108), $S_7$(009, 100), $S_8$(351, 158), $S_9$(101, 101), $S_{10}$(708, 109), $S_{11}$(669, 888), $S_{12}$(129, 104), $S_{13}$(344, 148), $S_{14}$(912, 108), $S_{15}$(001, 012), $S1_6$(366, 521), $S_{17}$(123, 149), and $S_{18}$(906, 612). The values of these nineteen audio samples are the following: $P_0$(0010111100111111), $S_1$(1110100110110011), $S_2$(1010000000001111), $S_3$(1110111111111111), $S_4$(11101001 00110000), $S_5$(1010101000000000), $S_6$(1111001011111111), $S_7$(1110100110110011), $S_8$(1010001100001111), $S_9$(1110111111111000), $S_{10}$(1110101100110010), $S_{11}$(1010101000000000), $S_{12}$(0011001011001111), $S_{13}$(0010100100110101), $S_{14}$(1010101000011001), $S_{15}$(0111001001101101), $S_{16}$(1010100100110100), $S_{17}$(0010101000100001), $S_{18}$(0111001011111000).
3. The chunks obtained in step 1.b are stored into the three LSBs of the audio samples selected in step 2. The results are as follows: $P_0$(0010111100111**110**), $S_1$(1110100110110**101**), $S_2$(1010000000001**111**), $S_3$(1110111111111**010**), $S_4$(11101001 00110**011**), $S_5$(1010101000000**101**), $S_6$(1111001011111**100**), $S_7$(1110100110110**110**), $S_8$(1010001100001**110**), $S_9$(1110111111111**001**), $S_{10}$(1110101100110**000**), $S_{11}$(1010101000000**001**), $S_{12}$(0011001011001**101**), $S_{13}$(0010100100110**010**), $S_{14}$(1010101000011**110**), $S_{15}$(0111001001101**111**), $S_{16}$(1010100100110**111**), $S_{17}$(0010101000100**001**), $S_{18}$(0111001011111**01**).

The bits marked in bold are chunks of the secret message that replaced the original three LSBs in the audio samples. The hiding capacity of the algorithm is equal to 3 bits out of 16 bits or 18% of the total size of the carrier file (3/16=0.18=18%). Figure 3 depicts a 1 minute audio segment from the original audio file before and after hiding the secret message D inside its samples.

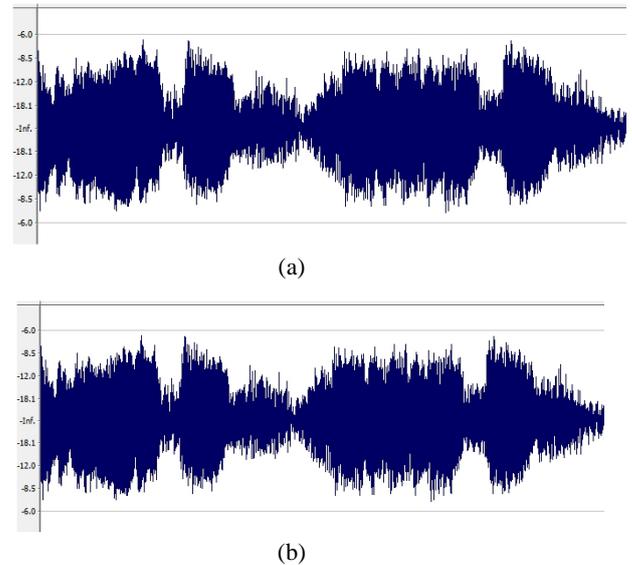

(a)

(b)

**Figure 3:** Before (a) and After (b) the covering process

4. The coordinates of the randomly selected audio samples are encoded into English sentences using the CFG and the lexicon of Figure 2 and Table 1. Since there are 19 samples, there should be 19 English sentences. They are as follows T={"those people study in the class", "the system sold in Georgia state", "sleep in the wood with tony", "Steve went rough on the teacher", "the dog barked on his friend", "drive on the road with caution", "move across the bridge with me", "Marc learn math in the college", "the boy went to high school", "a scholarship went to the students", "talk to the master with respect", "the house bought in Maryland county", "eat the apple in the kitchen", "John write circle on the wall", "dance in the night with mike", "a computer shipped to Atlanta city", "George does homework on the desk", "shoot the enemy by the sniper", "these guys dive in the ocean"}.
5. The final output is the carrier WAV audio file now carrying the secret data, and the English text T. Both are the intermediates to be sent to the intended receiver. It is worth noting that the algorithm does not guarantee that the generated English sentences are all semantically correct as it does not implement a semantic analyzer. It is the job of the user to compile the lexicon with the appropriate words that can semantically work with each other.

As for the uncovering process, it is the reverse of the above process. First, the second intermediate which is the English text T is decomposed into sentences such as T={$V_0,V_1,V_2,V_{t-1}$}. Then, each sentence is decomposed into words. Then, every word is mapped to a digit pertaining to the category in the lexicon it belongs to. At this point, coordinates of the carrier audio samples are generated and are used to extract the secret data from the three LSBs of the samples they point to. As a result, a string of bit is formulated which is then converted into ASCII text, revealing the original secret message D, mainly "kill joe".

## 6. CONCLUSIONS & FUTURE WORK

This paper proposed a steganography technique for hiding digital data into digital audio files based on a random algorithm to select the carrier audio samples into which bits of the secret data are to be hidden. It uses two intermediates to transmit the secret data: The first intermediate is a carrier audio file embedding the actual secret data in the three LSBs of its audio samples. The second intermediate is a well-structured text that encodes the random locations of the carrier audio samples in the carrier audio file. The key advantage of the proposed technique is the use of two intermediates that complement each other to convey the secret information. As a result, the proposed technique is less susceptible to stego-attacks as third parties often assume that the secret data are hidden in one intermediate and not in two intermediates that together are needed to decode the secret data. In that sense, the sender can first send one of the intermediates, and then later on, send the other one, misleading eavesdroppers from the true location of the secret data. Besides, the proposed technique has a second advantage; it is the random selection of audio samples to hide the secret data; thus, making it irrecoverable and hard for unauthorized third parties to predict the location of the secret data and recover them. Overall, the proposed technique exceptionally obscures data in such an anonymous way that the secret communication would pass undetected by forensics and passive eavesdroppers, and in case it was detected, it would be challenging then to uncover the secret data out of the communication.

Future research can improve upon the proposed technique so much so that it can take advantage of a semantic analyzer that would generate semantically structured English sentences using the presented context-free grammar. Consequently, the intermediate text would be even less suspicious and more trusting while being transmitted.

## ACKNOWLEDGMENTS

This research was funded by the Lebanese Association for Computational Sciences (LACSC), Beirut, Lebanon, under the "Stealthy Steganography Research Project – SSRP2012".